\documentstyle[aps,prb,multicol,graphicx]{revtex}

\begin{document}
\draft
\widetext

\title{Pressure Studies on a High-$T_c$ Superconductor Pseudogap
        and Critical Temperatures}

\author{E.V.L. de Mello$^1$, M.T.D. Orlando$^2$, J.L. Gonz\'alez$^1$,
E.S. Caixeiro and E. Baggio-Saitovich$^3$ }

\address{1-Departamento de F\'{\i}sica,
Universidade Federal Fluminense, Niter\'oi, RJ
24210-340, Brazil\\
2-Departamento de F\'{\i}sica, Universidade Federal do Espirito Santo,
ES 29060-900, Brazil\\
3-Centro Brasileiro de Pesquisas F\'isicas, Rio de Janeiro, RJ 22290-180 Brazil}
\date{Received \today }
\maketitle

\begin{abstract}

We report simultaneous hydrostatic pressure studies on the critical temperature 
$T_c$ and on the pseudogap temperature $T^*$ performed
through resistivity measurements  on an
optimally doped high-$T_c$ oxide 
$Hg_{0.82}Re_{0.18}Ba_2Ca_2Cu_3O_{8+\delta}$. 
The resistivity  is measured  as function of the temperature
for several different applied pressures around 1GPa.
We find that  $T^*$ increases linearly with the pressure 
at a larger rate than $T_c$. This result demonstrates
that the well known intrinsic pressure effect on $T_c$ 
is more important at the pair formation temperature  $T^*$.

\end{abstract}


\begin{multicols}{2}
\section{Introduction}
Immediately after the discovered of  high-$T_c$ superconductors\cite{BM}
high pressure experiments have played an important role in improving
and understanding these materials. However, despite of
tremendous efforts,
the  superconducting interaction 
in high-$T_c$  oxides\cite{BM}
remains one of the greatest 
puzzle in condensed matter physics.
More recently, several experiments\cite{TS} indicate that 
the appearance of a pseudogap, that is, a discrete structure
of the energy spectrum at a temperature $T^*$
above the superconducting phase transition at $T_c$ is
an important property which needs to be taken into account
in the quest for the solution of this problem\cite{Mello01}.
However, one  difficulty is that different techniques,
depending on the particular experimental probe used, 
may yield different  values of  pseudogap temperature $T^*$ 
for the same type of sample.

There are  mounting evidences that the pseudogap
and the charge inhomogeneities, possibly in a stripe
morphology\cite{Tranquada,Bianconi}, are different  but
closely related phenomena.
This is mainly because the $T^*$ has its maximum
value for underdoped compounds\cite{TS} which possess the larger
charge inhomogeneities\cite{Zhou,Setal99}. Furthermore $T^*$ decreases with
the average doping level and probably becomes equal
to $T_c$ for overdoped  compounds which are those with
more homogeneous charge distributions\cite{Bozin,Egami}.
Moreover, a local Meissner state, which
usually appears only in the superconducting phase,
has been detected far above $T_c$ for an underdoped sample\cite{IYS}. Such
inhomogeneous diamagnetic domains develop
near\cite{Oda}  $T^*$ for $La_{2-x}Sr_xCuO4$ thin films and grow continuously  
as  the temperature decreases towards $T_c$.
Near $T_c$, the domains appear to percolate through the sample.

Based on these  experimental findings, 
some of us  have developed a new approach to deal with  this
phenomenology\cite{Mello01}. The main ideas are the following; 
compounds of a given family, with  an average charge density $\rho_m$
have  inhomogeneous charge distributions which are less homogeneous
for underdoped and more homogeneous for overdoped samples. These 
distributions contain a hole-rich 
and a hole-poor partition which mimic the stripes. These
charge inhomogeneities in the hole-rich regions produce local clusters with
spatially dependent
superconducting gaps $\Delta_{sc}(r)$ due to local
Cooper pairing and  superconducting
temperatures $T_c(r)$. 
$T^*$ is the largest of all the $T_c(r)$, that is, the largest local
superconducting temperature in a given compound. Since the clusters
have different charge densities, some are metallic and some are insulating.
As the temperature decreases below the $T^*$, part of the metallic clusters 
become superconducting and, eventually a temperature is reached
where the different superconducting regions can percolate, exactly as 
clearly demonstrated by the fine 
diamagnetic domains measured by Iguchi et al\cite{IYS}. At or
below the percolation temperature, the system can hold a 
dissipationless supercurrent and this temperature is normally 
identified as superconducting critical temperature $T_c$. Thus,
the temperature $T_c$ is assumed to be the percolating transition for 
the supercurrent and it is not the usual superconducting transition
related with the appearance of a superconducting gap. This 
scenario resembles the proposal of Ovchinnikov et al\cite{OWK} and its 
implications are discussed in Ref.3.

According to these ideas, experiments which produce changes on $T^*$ 
are able to provide direct information on the superconducting interaction.
One such type of experiment is the isotopic substitution of $^{16}O$
by $^{18}O$ in the slightly underdoped $HoBa_2Cu_4O_8$ which leads
to an increase in $T^*$ from $170$K to $220$K and negligible 
effect on $T_c$\cite{Temprano}. Despite the anomalous increase of
$T^*$ for the heavier isotope, this result indicated that, at least
for this sample, electron-phonon induced effects are likely
to be present in the mechanism related to $T^*$ and somehow, are
less important on $T_c$.

The prime objective of this work is to study the nature of the pseudogap
and to achieve this goal we  analyze resistivity measurements
under low applied hydrostatic
pressures.  The measurements were performed on
two different  optimally doped 
$Hg_{0.82}Re_{0.18}Ba_2Ca_2Cu_3O_{8+\delta}$ samples in
temperatures range from 120K (below $T_c$)  up to room temperatures.
The resistivity for underdoped and optimally doped high-$T_c$ oxides
has a linear behavior up to very high temperatures but  at $T^*$, it downturns
and falls faster as temperature is decreasing\cite{TS}. 
We have recently reported
pressure effects  on $T_c$ on this compound\cite{Orlando,Jorge},
but our aim here is to study simultaneous pressure effects on $T^*$ and $T_c$. 
We show that  $T^*$ increases at larger rate than $T_c$ 
in the pressure range of our experiment.
This result provides a novel interpretation for the well known
{\it pressure intrinsic effect} and we argue that the data are 
consistent with a superconducting interaction mediated
by phonons.

\section{Experimental Measurements}

The resistivity measurement were performed using an hydrostatic pressure
cell within a cooper cylinder piston. An external pressure was applied to
the cylinder and it was transmitted through a n-pentane-isoamyl alcohol
mixture (1:1) to the sample. The sample was mounted on a Teflon sample
holder, and in order to provide a good thermal contact with the temperature
sensor, the sample holder was thermally coupled to the CuBe pressure
cell. The
pressure was monitored with a manganin resistance situated inside the cell.
Further details of this experimental set-up can be found 
elsewhere\cite{Orlando,Jorge,Fernandes,Thompson}.

Four electric contacts with low electric resistance were deposited onto the
sample by using silver paint. After stabilizing the pressure,
the temperature was increase recording the resistance on the sample. This
procedure was repeated at several pressures making $R(T,P)$ curves. The
temperature was recorded and stabilized using a "Lake-Shore-340"
temperature controller. The resistance
measurements were performed using an AC-bridge resistance "Linear Research
Inc, model LR-700". The $R(T,P)$ curves were measured within the
linear-response regime with an applied current to the sample of 
$10 \mu$A.

\section{Results and Discussion}

We have measured the resistivity for two optimally doped
$Hg_{0.82}Re_{0.18}Ba_2Ca_2Cu_3O_{8+\delta}$ samples (A and B)
as function of temperature for
four different applied pressures up to $1.23$GPa. The samples
were made under the same conditions.
As an example of our results, the  temperature dependent resistivity 
data for sample A at zero pressure and $12.3$kbar is shown in Fig.1. 

Panel 1a shows the data for $R(T)$ and 1b shows a plot of
$(R(T)-R_0)/\alpha T$ against T, where $R_0$ and $\alpha$ 
are respectively, the 
intercept and the linear coefficient of the high temperature linear
part of the $R(T)$ curve. 
The $T^*$ temperature was taken at the point where the experimental resistance
curve starts to downturn from the high-temperature linear behavior. 
To determine $T^*$, we have used the same
criterion of Ito et al\cite{Ito} which is shown in Fig.1b.
At each applied pressure, $T^*$ 
was taken as the temperature
at which the curve starts to downturns from its base-value as
demonstrated in panels 1b and 1d.
It should be noted that others criteria for $T^*$ leaded to
practically the same experimental findings, meaning that
the $T^*$  increase with the pressure is independent of its
experimental criterion.
\begin{figure}
\includegraphics[width=8.5cm]{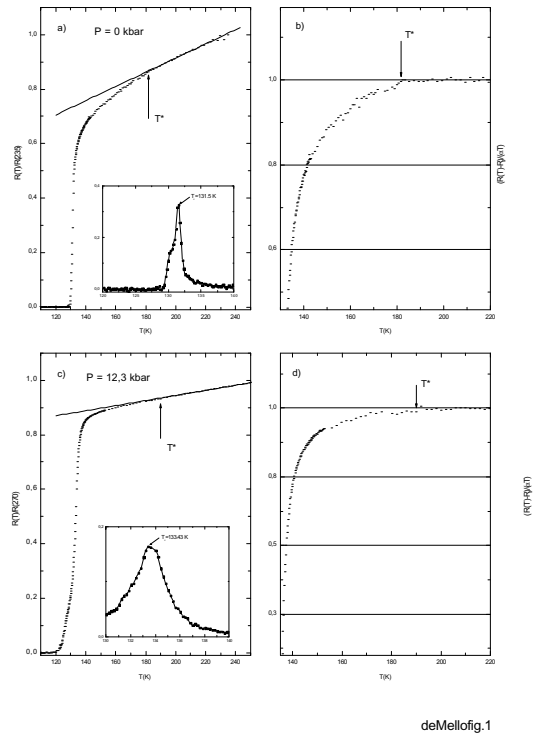}
\caption{The resistivity measurements for sample A as function 
of the temperature for
zero and 12.3 Kbar. The criterion for  $T_c$ (where $dR/dT$ has the 
largest value) is shown in the insert of
panels a) and c). The criterion for  $T^*$ is shown in panels b) and d).}
\end{figure}

The critical temperature $T_c$ was defined as the
temperature for which the $dR/dT$ has the largest value\cite{Jover} 
as shown in panels 1a and 1c. We have measured an experimental $dT_c/dP$
rate of 0.15 K/kbar for sample A and 0.26K/kbar for sample B, which 
are typical values for these multilayer
compounds\cite{Jorge}. In order to verified these derivatives we have
performed  magnetic susceptibility  measurements on both samples.
The increase of $T_c$ with the applied pressure is in the
range of similar compounds which we have
previously been reported\cite{Orlando,Jorge}. While sample A has
$T_c\approx 131.5$K and $T^* \approx 180$K at ambient pressure, 
sample B has $T_c\approx 132.0$K and $T^*$
just above $160$K. Since both the present samples and the ones
previously studied were made  under the same condition, we do not have
an explanation for these differences. Nevertheless it is a 
common fact that different samples of the same compound with
the same $T_c$, may
exhibit large differences in their $T^*$ value\cite{TS}.

The theory of pressure effects\cite{Taka,Wijngaarden} 
identify two contributions
to the changes on $T_c$: a pressure induce charge transfer (PICT)
from the charge reservoirs to
the $CuO_2$ planes and an intrinsic term, of unknown origin, 
which is more clearly detected mainly
at, or near to, the optimally doped compounds. In the case of optimal
doping and low pressure, i.e., less than 5GPa,  there is a
negligible PICT\cite{Orlando,Wijngaarden} and  $T_c$  is a linear
function of the pressure. This linear behavior is therefore
attributed entirely to the intrinsic contribution\cite{Taka,Wijngaarden}
and it is believed  to be due to the effect of the
pressure on the superconducting interaction\cite{Mello97,Angi}. 
It  can be regarded as due to
an intrinsic  increase on the superconducting interaction, what makes 
the Cooper pairs  more tightly bound and also increases the
superconducting gap.

Thus, in order to determine whether the pseudogap is also affected by
an external pressure and if the intrinsic effect is also present
at the pseudogap temperatures, we 
plotted in Fig.2a  and 2b the $T_c$  and  $T^*$ values obtained 
from the resistivity data.  While sample A has $T^* \approx 000$K
at ambient pressure, sample B hat $T^* \approx 138$K. Through
these figures we can 
see for the first time, the original and most
important result of our work, that  $T^*$ and $T_c$ increase 
under the applied pressure in a linear way, but $dT^*/dP$ is
around 3-5 times larger than $dT_c/dP$. For sample A we get 
$dT^*/dP=0.75$K/kbar and $dT^*/dP=0.63$K/kbar for sample B.
This behavior lead us to the main conclusion
of this work, namely,  the 
pressure induced intrinsic term modifying $T_c$ is  
much stronger on the pairing formation temperature $T^*$.
\begin{figure}
\includegraphics[width=8.5cm]{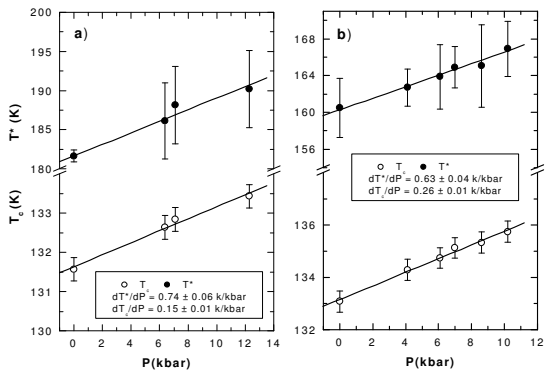}
\caption{The estimated values for $T_c$ and $T^*$ as function of the
applied pressure for both samples . 
Data in panel a) is for sample B and b) for sample B.
Notice that both samples have   linear variations pressure
variations for $T_c$  and  $T^*$.}
\end{figure}

Our  experiment  reveals a larger pressure effect on  $T^*$ than
on $T_c$. We have already mentioned the results of Temprano et
al\cite{Temprano} with  different isotope effect on these
temperatures. Similarly Harris et al\cite{Harris} have demonstrated
with ARPES measurements that  the zero temperature 
superconducting gap $\Delta_{sc}$ does not correlates with $T_c$.
However, according to the percolation scenario
for high-$T_c$ oxides\cite{Mello01},  $\Delta_{sc}$ 
correlates with the onset of superconducting gap $T^*$ and 
not with  $T_c$ (see Fig.3 of Ref.3).

We have mentioned that our most important finding is the
larger linear increase on $T^*$ than on $T_c$  and
such behavior is possibly the consequence  of the intrinsic pressure effect 
on the superconducting interaction\cite{Mello97,Angi}.
The presence of a constant intrinsic term on both $T_c$ and $T^*$ 
is an indirect evidence of the increase  of the superconducting interaction.
However, we will argue that this result could either be explained under the
assumption of a phonon mediated mechanism or by a non-equivalent
two layer hole density.
%
In a typical BCS superconductor, either in weak\cite{deGennes} or 
strong\cite{Carbotte} coupling regime, $T_c$ is proportional 
to the Debye frequency $\hbar \omega_D$ and to 
$exp(-1/N(0)V)$, where $N(0)$ is the density of state and 
$V$ the phonon mediated pairing potential amplitude. 
These two terms yield opposite contributions when the sample
is under an external applied pressure. The applied
pressure increases $\omega_D$ in any solid and also broaden the
density of states, thereby reducing $N(0)$ which is the
dominant factor in a typical BCS superconductor and produces an
overall decrease in  $T_c$. Now, assuming the same type of
dependence on high-$T_c$ superconductors, the effect on 
$N(0)$ is minimized in
a high-$T_c$ sample with a large charge inhomogeneities 
distribution because such inhomogeneities  produce also a broadening of the
density of states, as clearly demonstrated by
Ovchinnikov et al\cite{OWK}. Therefore the net low pressure effect 
on an optimally doped cuprate is
to increase $\omega_D$  which yields a proportional  linear increase 
on all $T_c(r)$ in the sample  and, consequently, on $T^*$, 
assuming, according Ref.3, that $T^*$ is the
pair formation temperature. Since at the low pressures
all values of $T_c(r)$ throughout the sample increase 
proportionally to $\omega_D$, 
the percolation temperature $T_c$ will
also increases. This provides  a physical interpretation
to the linear behavior of both $T^*$ and $T_c$ and to the well known  
pressure intrinsic effect which, despite of being widely
detected\cite{Taka,Wijngaarden,Mello97,Angi}, has not received any previous
microscopic interpretation.

 Another interesting possibility to explain the increase of  $T_c$
and $T^*$ under pressure, uses the idea of a  competition between
two hole dependent effects\cite{Kivelson}: Pairing formation which
decreases with doping and  
the system stiffness to phase fluctuations which controls long-range
phase coherence and increases with doping. On the other hand, 
it is well known that three layer compounds, like our 
$Hg_{0.82}Re_{0.18}Ba_2Ca_2Cu_3O_{8+\delta}$ samples A and B,  
must have differences 
in the doped hole concentration in the different 
layers\cite{Jover,Wijngaarden}. Therefore one  layer may have
a hole concentration with favor pairing and the other may favor 
phase coherence. The pressure  induced charge transfer (PICT) 
between different layers may lead to an optimal  dependent 
distribution of doped 
holes which increases the layer superfluid  stiffness, producing an
enhancement in $T_c$, concomitantly with  an increase in the 
pairing scale, and therefore an enhancement in $T^*$\cite{Kivelson}.

\section{Conclusion}

We have reported here for the first time simultaneously  hydrostatic pressure
effects on the pseudogap temperature $T^*$ and on
$T_c$ of an optimally doped high-$T_c$
superconductor. 
We have measured a larger linear increase on $T^*$
than on $T_c$ under  applied pressures up to $12.3$GPa. 
These data show that the  intrinsic effect on $T_c$
could be a consequence of the larger  effect on 
the pair formation temperature $T^*$.

There are several implications from the  pressure induced increase
of $T^*$ and $T_c$. The  increase on  $T^*$  
indicates the presence of the   superconducting mechanism 
what supports the local Cooper pairing scenario due to
the charge inhomogeneities in the planes of $CuO_2$. 

Moreover, the linear behavior of  $T^*$ and $T_c$ under pressure
is consistent with a
mechanism leading to the formation of superconductivity
mediated by phonons. This is  in agreement with
the very large isotope effect measured on $T^*$ by
Temprano et al\cite{Temprano} and very recent ARPES
measurements\cite{Lanzara}. These experiments
together yield a strong support for the lattice
effects on the mechanism for high-$T_c$ superconductivity. 
Another interesting implication, also consequence of a
phonon mediated superconducting interaction and the charge inhomogeneities, 
is a novel physical interpretation for the
well known pressure intrinsic effect on $T^*$
as due to the pressure change of the Debye frequency. However
the enhancement of $T^*$ and $T_c$ may also be due to the 
PICT between non-equivalent layers of our sample.

We are presently pursuing our resistivity
under pressure experiment
with  samples having different oxygen content. 
We believe that for underdoped compounds,
$T^*$ and $T_c$ will behave in similar way as reported above.
For more homogeneous overdoped samples we 
expect the broadening of
$N(0)$ under applied pressure to be much less important 
and therefore $T^*$ and $T_c$ may not increase. At higher
pressures, for instance above 5GPa, and different doping
level, we expected the PICT
to be important leading to different
pressure variations of $T^*$ and $T_c$.

Financial support from Pronex, CNPq and FAPERJ is gratefully acknowledged.
JLG thanks CLAF for a CLAF-CNPq pos-doctoral fellowship.

\end{multicols}
\end{document}